\documentclass[12pt]{article}

\usepackage{graphicx}	
\usepackage{amsmath}	
\usepackage{amssymb}	
\usepackage{xcolor}
\usepackage{microtype}
\usepackage[a4paper, margin=2.4cm]{geometry}
\usepackage{url}

\usepackage{mathptmx}

\usepackage{helvet}

\usepackage[font=sf]{caption}

\usepackage{sectsty}
\allsectionsfont{\bfseries\sffamily}

\def\M{M$_\odot$}
\def\ergs{erg\,s$^{-1}$}
\def\kms{km\,s$^{-1}$}

\makeatletter
\renewcommand{\maketitle}{\bgroup\setlength{\parindent}{0pt}
\begin{flushleft}
  \textbf{\LARGE \@title}
  \\
  \vspace{0.5cm}
  \@author
\end{flushleft}\egroup
}
\makeatother

\title{An extremely energetic supernova from a very massive star in a dense medium}

\date{}

\author{\large{Matt Nicholl,$^{1,2,*}$
Peter K.~Blanchard,$^{3,4}$
Edo Berger,$^{4}$
Ryan Chornock,$^{5}$ Raffaella Margutti,$^{3}$
Sebastian Gomez,$^{4}$
Ragnhild Lunnan,$^{6,7}$
Adam A. Miller,$^{3,8}$
Wen-fai Fong,$^{3}$
Giacomo Terreran,$^{3}$
Alejandro Vigna-G\'omez$^{9}$,
Kornpob Bhirombhakdi$^{10}$,
Allyson Bieryla$^{4}$,
Pete Challis$^{4}$,
Russ R. Laher$^{11}$,
Frank J.~Masci$^{11}$ \&
Kerry Paterson$^{3}$}
\\
  \vspace{0.5cm}
\normalsize
$^{1}$Birmingham Institute for Gravitational Wave Astronomy and School of Physics and Astronomy, University of Birmingham, Birmingham B15 2TT, UK \\
$^{2}$Institute for Astronomy, University of Edinburgh, Royal Observatory, Blackford Hill, EH9 3HJ, UK \\
$^{3}$Center for Interdisciplinary Exploration and Research in Astrophysics (CIERA) and Department of Physics and Astronomy, Northwestern University, Evanston, IL 60208, USA \\
$^{4}$Center for Astrophysics  $\vert$ Harvard \& Smithsonian, 60 Garden Street, Cambridge, MA, 02138, USA \\
$^{5}$Astrophysical Institute, Department of Physics and Astronomy, 251B Clippinger Lab, Ohio University, Athens, OH, 45701, USA \\
$^{6}$Oskar Klein Centre, Department of Astronomy, Stockholm University, Stockholm, Sweden \\
$^{7}$Department of Astronomy, California Institute of Technology, Pasadena, CA, USA \\
$^{8}$The Adler Planetarium, Chicago, IL, 60605, USA \\
$^{9}$DARK, Niels Bohr Institute, University of Copenhagen, Blegdamsvej 17, 2100, Copenhagen, Denmark \\
$^{10}$Space Telescope Science Institute, 3700 San Martin Drive, Baltimore, MD 21218, USA \\
$^{11}$IPAC, California Institute of Technology, 1200 E. California Blvd, Pasadena, CA 91125, USA \\
$^*$mnicholl@star.sr.bham.ac.uk
}

\begin{document}
\maketitle

\noindent\textbf{
The interaction of a supernova (SN) with a circumstellar medium (CSM) can dramatically increase the emitted luminosity by converting kinetic energy to thermal energy. In `superluminous' supernovae (SLSNe) of Type IIn -- named for narrow hydrogen lines in their spectra \cite{schlegel1990} -- the integrated emission can reach $\sim 10^{51}$\,erg \cite{smith2007a,drake2010,chatzopoulos2011,rest2011,benetti2014}, attainable by thermalising most of the kinetic energy of a conventional SN. A few transients in the centres of active galaxies have shown similar spectra and even larger energies \cite{blanchard2017,kankare2017}, but are difficult to distinguish from accretion onto the supermassive black hole. Here we present a new event, SN2016aps, offset from the centre of a low-mass galaxy, that radiated $\gtrsim5\times10^{51}$\,erg, necessitating a hyper-energetic supernova explosion. We find a total (SN ejecta $+$ CSM) mass likely exceeding $50-100$\,\M, with energy $\gtrsim 10^{52}$\,erg, consistent with some models of pair-instability supernovae (PISNe) or pulsational PISNe -- theoretically-predicted thermonuclear explosions from helium cores $>50$\,\M. Independent of the explosion mechanism, this event demonstrates the existence of extremely energetic stellar explosions, detectable at very high redshifts, and provides insight into dense CSM formation in the most massive stars.
}

\medskip

\begin{figure*}
    \centering
    \includegraphics[width=12cm]{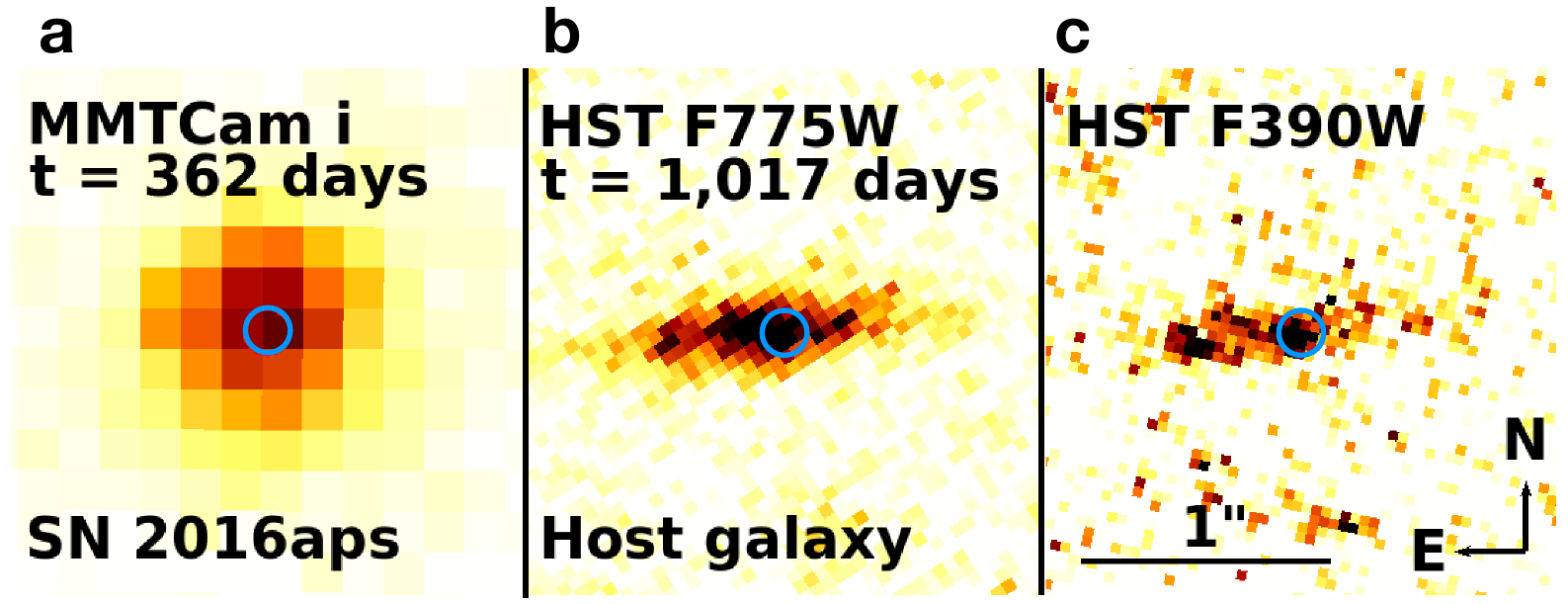}
    \caption{
    Ground-based and \textit{Hubble Space Telescope} images of SN2016aps and its host galaxy. (a) MMTCam $i$-band ($7730$\,\AA) image of SN2016aps at a phase of 362 days. (b) \textit{HST} optical F775W-band image of the host galaxy obtained at a phase of 1017 days. (c) \textit{HST} UV F390W-band image of the host galaxy at the same phase. All images were astrometically aligned using a catalog of matched sources in the field of view (not shown at this scale). The blue circle shows the position of SN2016aps, as measured in the MMTCam image, with its radius equal to the combined $3\sigma$ error from determining the centroid and aligning the MMT image to the \textit{HST} images (Methods). SN2016aps is coincident with the brightest UV emitting region of its host galaxy, providing a direct connection with on-going star formation activity.  The location of SN2016aps is $0.15''$ from the optical center of its host galaxy. }
    \label{fig:hst}
\end{figure*}
\begin{figure*}[ht]
    \centering
    \includegraphics[width=16cm]{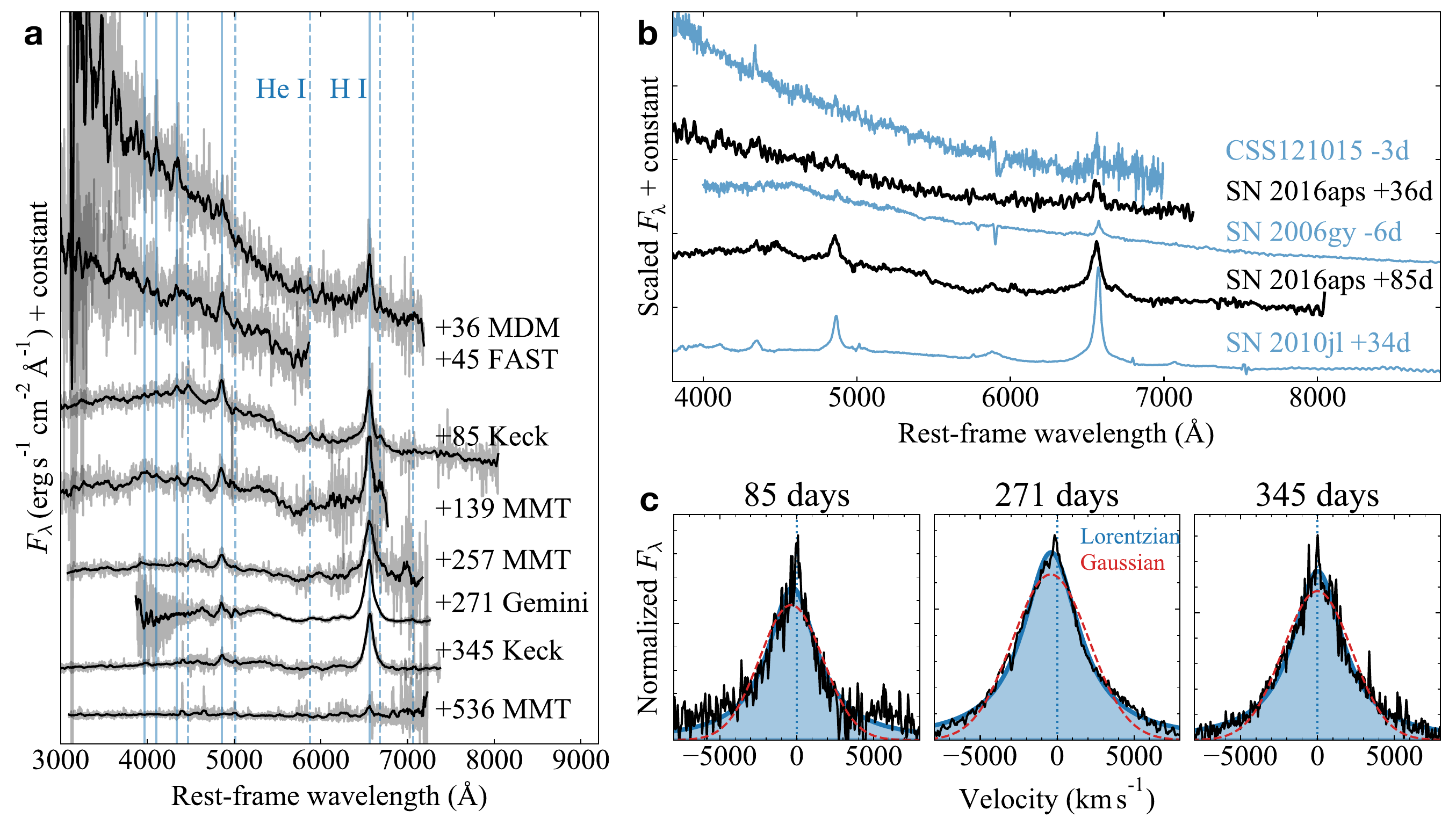}
    \caption{
    Spectroscopic evolution of SN2016aps over 500 rest-frame days following discovery. (a) The spectra have been smoothed using a Savitsky-Golay filter, with lighter colours showing the original unsmoothed data, and offset for clarity. Phases (days since maximum brightness) are labelled in the SN rest frame, based on a redshift of $z=0.2657$. Vertical lines mark the dominant emission lines from neutral hydrogen and helium. (b) Comparison of two representative spectra of SN2016aps to other SLSNe IIn \cite{smith2007a,benetti2014,fransson2014}. SN2016aps shows an early blue continuum and Balmer lines that increase in equivalent width over time, similar to previous events. (c) Continuum-subtracted H$\alpha$ emission in the high signal-to-noise ratio (S/N) spectra from Gemini and Keck.  Gaussian and Lorentzian functions have been fitted to the line profiles, with the latter better reproducing the narrow core and broad wings. The velocity full-width at half maximum is $4090\pm 230$\,\kms\ (1$\sigma$) at $80-350$ days, indicating an optical depth $\tau\gtrsim6$ to H$\alpha$ photons \cite{fransson2014}.}
    \label{fig:spec}
\end{figure*}
\begin{figure*}[t]
\centering
    \includegraphics[width=12cm]{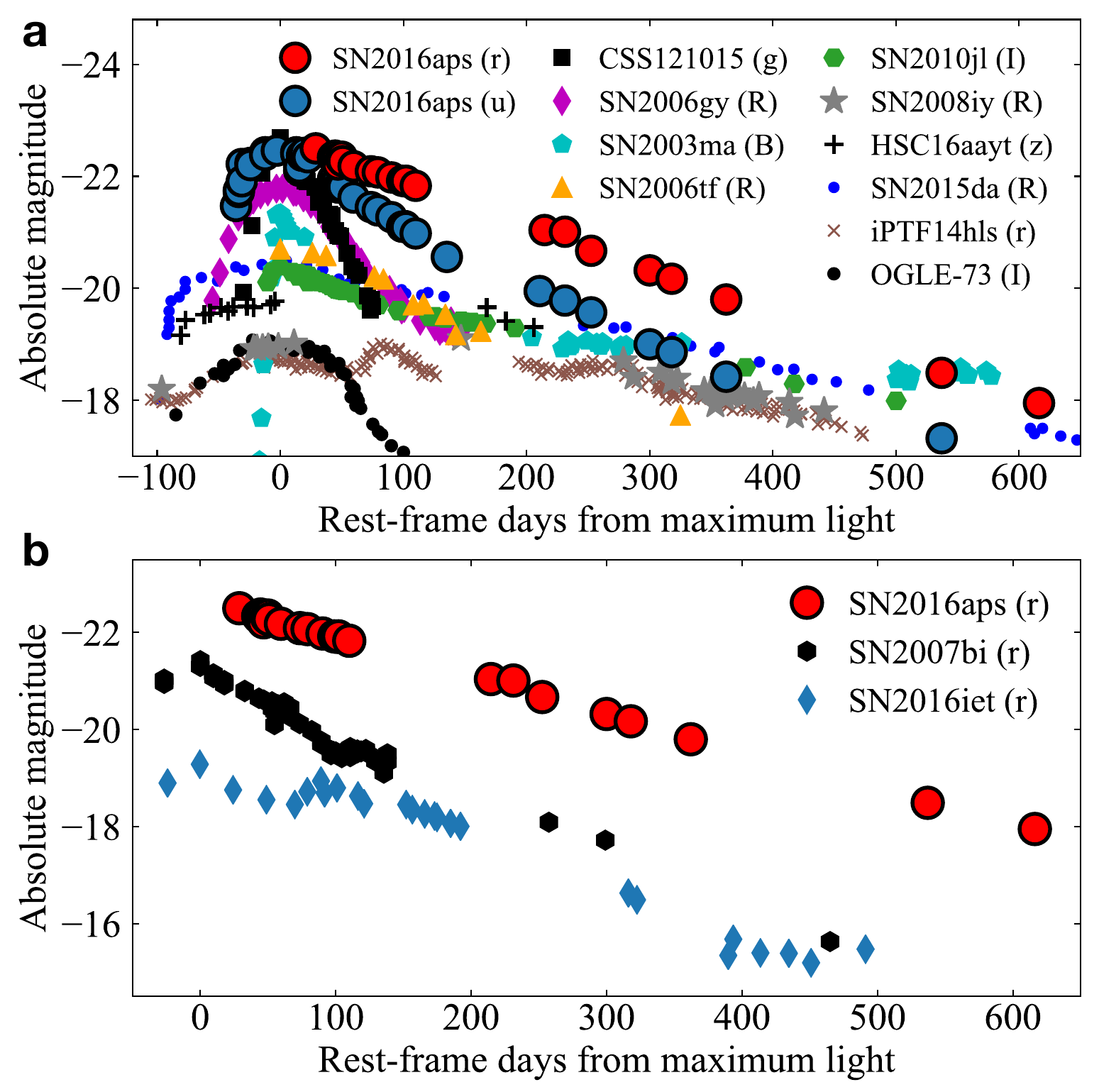}
    \caption{
    Optical light curve of SN2016aps, in comparison to previous energetic SNe. (a) Light curve of SN2016aps in rest-frame $r$ ($6260$\,\AA) and $u$ bands ($3560$\,\AA) compared to other luminous and/or long-lived SNe II/IIn on an absolute magnitude scale (Methods). The available band closest to rest-frame $r$ was chosen for each comparison SN. SN2016aps has a peak brightness at least as bright as any other SLSN IIn, but a slow decline rate (0.78\,mag per 100 days in rest-frame $r$ band) previously only seen in lower-luminosity events, resulting in an integrated electromagnetic output several times greater. (b) Comparison of SN2016aps to PISN and PPISN candidates, SN2007bi \cite{galyam2009b} and SN2016iet \cite{gomez2019}. SN2016aps clearly exceeds both the peak luminosity and radiated energy of these previous events.}
    \label{fig:lc}
\end{figure*}
\begin{figure*}
    \centering
    \includegraphics[width=13cm]{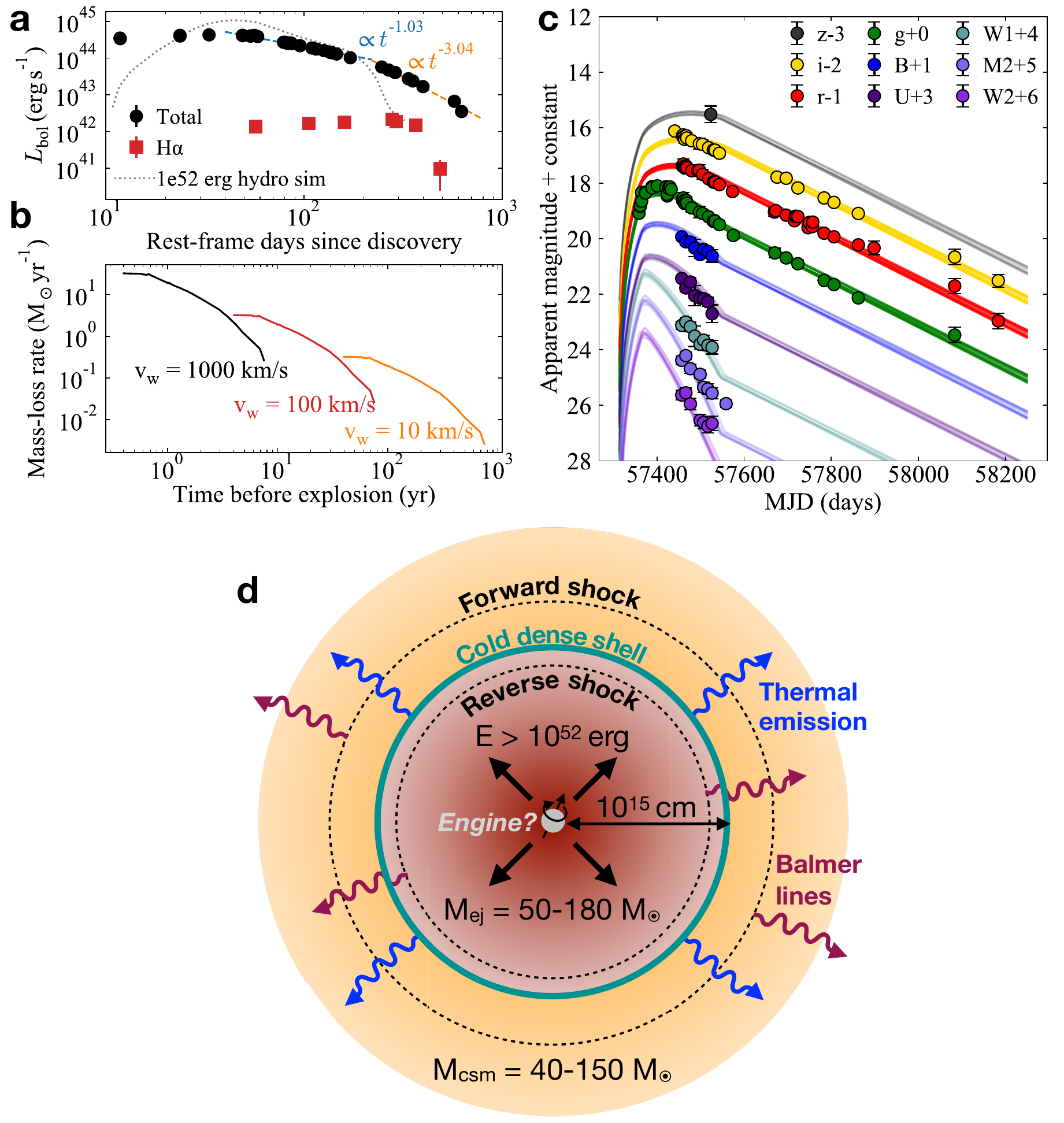}
    \caption{
    Full light curves of SN2016aps and derived properties. (a) Bolometric light curve and H$\alpha$ luminosity. The data can be fitted with a broken power law transitioning from a shallow to a steep decline; the H$\alpha$ luminosity drops steeply $\sim100$ days later. These drops may indicate the end of shock heating and a transition to diffusion-dominated luminosity \cite{fransson2014,smith2010}. Overplotted is a SLSN IIn radiation hydrodynamics model for a CSM mass of 17.3\,\M\ and explosion energy of $10^{52}$\,erg \cite{dessart2015}. Although it can reproduce the peak luminosity, a larger mass is needed to match the long timescale of SN2016aps. All error bars show $1\sigma$ uncertainties. (b) Mass-loss history inferred from the bolometric light curve for different assumed wind velocities \cite{chugai1994}. The progenitor of SN2016aps requires a mass-loss rate up to 6 orders of magnitude greater than expected for blue supergiant winds \cite{vink2018}. (c) Multi-colour light curves in optical and UV bands. We overlay realizations of a CSM interaction model \cite{chatzopoulos2012} from \textsc{mosfit} \cite{guillochon2018}; the best fit has $M_{\rm ej}\sim M_{\rm CSM}\gtrsim150$\,\M, and $E_K\approx3\times10^{52}$\,erg. Acceptable fits are possible with $M_{\rm ej}\gtrsim50$\,\M, $M_{\rm CSM}\gtrsim40$\,\M, though these under-predict the UV bands (Methods). (d) Schematic illustrating the model and inferred parameters. For ejecta (core) masses $<65$\,\M\ (below the PISN threshold), a magnetar engine may be needed to increase the kinetic energy or luminosity.
}
    \label{fig:bol}
\end{figure*}

\noindent SN2016aps (internal designation, PS16aqy) was discovered by the Pan-STARRS Survey for Transients \cite{huber2015} on 2016 February 22 UT with an apparent magnitude $m=18.12\pm0.08$\,mag in the $i$ band (7545\,\AA). We selected this target for spectroscopic follow-up due to its large brightness contrast relative to the previously undetected host galaxy, with $m\gtrsim 23.5$ mag in Pan-STARRS1 $3\pi$ survey data.  Our first spectrum \cite{chornock2016}, on 2016 March 2 UT, showed hydrogen Balmer emission lines at a redshift of $z=0.2657$, and hence an absolute magnitude of $M=-22.5\pm 0.08$\,mag at the time of discovery (Methods). A search of archival images from the intermediate Palomar Transient Factory \cite{law2009} revealed a rising light curve going back to at least 2015 December 02 UT, with maximum brightness on 2016 January 17 UT. We obtained further spectra spanning 500 days, and optical and UV imaging spanning 1000 days. All phases are in the SN rest-frame relative to the date of maximum brightness.\\
\\
We imaged the location of SN2016aps 1,017 rest-frame days after maximum brightness using the \textit{Hubble Space Telescope}, identifying the previously undetected host galaxy (Figure \ref{fig:hst}). The absolute magnitude of the galaxy, $M_{F775W}=-17$\,mag (intermediate between the Small and Large Magellanic Clouds), indicates a stellar mass $M_*\sim10^8$\,\M\ (assuming $M_*/M_\odot\sim L_*/L_\odot$) and likely a sub-solar metallicity (Methods). A compact bright region, visible in both the F775W (optical) and F390W (ultraviolet) images, is coincident with the SN position and significantly offset from the center of the galaxy by $0.15''\pm 0.03''$ (Methods). While some energetic, hydrogen-rich transients may represent supermassive black hole accretion rather than SNe \cite{blanchard2017,kankare2017,dong2016}, the faint galaxy and offset from the nucleus require a SN origin for SN2016aps. The UV image shows that it occurred in the brightest star-forming region within the galaxy, pointing to a massive star progenitor.\\
\\
The spectra (Figure \ref{fig:spec}) are typical of SLSNe IIn \cite{smith2007a,benetti2014}, while the peak absolute magnitude is equal to the most luminous events \cite{benetti2014} (Figure \ref{fig:lc}). What sets SN2016aps apart from all previous events is the unprecedented combination of a large peak luminosity characteristic of the brightest `compact shell' SLSNe IIn \cite{smith2007a}, and a slow rate of fading (0.8 mag per 100 days) similar to long-lived events \cite{smith2008,fransson2014}, thought to have more extended CSM. To measure the total optical output of SN2016aps, we integrate our photometry at each point in the light curve, and fit a blackbody function to estimate the flux outside of the observed bands (Methods). The radius is roughly constant at $5\times 10^{15}$\,cm, while the temperature decreases from 10,000\,K near peak to 6,000\,K after 200 days (Extended Data Figure 1). The peak luminosity is $4.3\times 10^{44}$\,\ergs, and the integrated energy radiated over the time of our observations is $E_{\rm rad}=(5.0\pm 0.1)\times10^{51}$\,erg. This is the largest radiated energy for any confirmed SN; the maximum observed energy for previous SLSNe is $\approx 2\times 10^{51}$\,erg \cite{smith2007a,drake2010,chatzopoulos2011,rest2011,benetti2014}. It was argued that SN2003ma may have emitted up to $4\times10^{51}$\,erg \cite{rest2011}, but this is highly uncertain as only 20\% of this energy was within the wavelength range covered by observations, compared to 70\% for SN2016aps. Where a normal SN has (just) enough kinetic energy to power previous SLSNe IIn, the total emission from SN2016aps cannot be explained without a hyper-energetic explosion \cite{sukhbold2016}.\\
\\
Assuming the light curve is powered by shock-heating of CSM, we use common relations \cite{chevalier2011} to estimate the kinetic energy and shock velocity from the luminosity and total emission (see Supplementary Information). We find an explosion energy $E_K^2/M_{\rm ej}=4.9$ (in units of $10^{51}$ erg and \M) and a shock velocity of $v_s\approx 4,600$\,\kms. In this model, $E_{\rm rad}/E_K=0.32$ \cite{chevalier2011}, giving an ejected mass $M_{\rm ej}\gtrsim 52$\,\M\ for our measured $E_{\rm rad}=5\times10^{51}$\,erg. We map the pre-explosion mass-loss, $\dot{M}$, by inverting the bolometric light curve according to $L=\frac{1}{2}\psi\dot{M}v_s^3/v_w$ \cite{chugai1994}, where $v_s$ is the shock velocity and $\psi\sim 0.5$ is the efficiency of converting kinetic energy to radiation \cite{chevalier2012}. This gives $\dot{M}\sim 0.1-10$\,\M\,yr$^{-1}$ for a wind with velocity $v_w=10-1{,}000$\,\kms\ (Figure \ref{fig:bol}). We can estimate the time of mass ejection as $R_s/v_w$, where the shock radius $R_s=v_s(t-t_{\rm explosion})$. The integrated CSM mass, lost years to centuries before explosion, is $M_{\rm CSM}\gtrsim40$\,\M. Photoionization from external UV radiation in the local star-forming region may help to confine this mass-loss close to the progenitor \cite{mackey2014}. However, compared to a blue supergiant star with $v_w\sim 1{,}000$\,\kms, the inferred mass loss rate exceeds typical values by 6 orders of magnitude \cite{vink2018}.\\
\\
As the CSM must be ejected shortly before explosion, a constant density corresponding to a single massive outburst may be more appropriate than a wind. We use the CSM model \cite{chatzopoulos2012,guillochon2018} in the light curve fitting package \textsc{mosfit} to fit the multi-band data (Supplementary Information). Although we can reasonably reproduce the light curve with the parameters estimated above, it underestimates the UV luminosity (Extended Data Figure 2). If we vary these parameters in a Markov Chain Monte Carlo fit, we find that the formal best fit (Figure \ref{fig:bol}) has $M_{\rm ej}=182^{+42}_{-23}$\,\M, $v_{\rm ej}=6{,}015^{+134}_{-134}$\,\kms, and $M_{\rm CSM}=158^{+23}_{-20}$\,\M\ (uncertainties correspond to $1\sigma$ range). These posteriors are relatively insensitive to the CSM density profile (Extended Data Figures 2, 3).
The estimated masses should be treated with caution due to various simplifications (central heat input, constant opacity, neglecting recombination) inherent in modelling a complex process with an analytic formalism. However, the requirement for $\gtrsim{\rm few}\times10$\,\M\ of ejecta and CSM is robust, as evidenced by the long timescale of the light curve and optically thick spectrum, and comparison to more detailed hydrodynamic models \cite{dessart2015} (Figure \ref{fig:bol}). We now turn to progenitor and explosion scenarios that can explain the extreme radiated energy in combination with a massive and hydrogen-rich CSM.\\
\\
Stars with initial masses of $70-140$\,\M\ experience large `pulsational PISN' (PPISN) eruptions following core carbon burning \cite{woosley2017}, before an eventual iron core collapse. A recent example may have been observed in the hydrogen-poor SN2016iet \cite{gomez2019}. Specifically, a progenitor with a helium core $\simeq40-50$\,\M\ (total initial mass $\approx100$\,\M\ at SMC metallicity) begins pulsing and ejects its envelope $\sim$years before core collapse. Single star models  have difficulty retaining hydrogen rich material until the final years before explosion, but mergers in massive binaries can produce the same core mass while retaining a hydrogen envelope \cite{vignagomez2019}. The rate of mergers in the necessary mass range is estimated to be $\sim 1.4\times10^{-3}$ of the core-collapse SN rate \cite{vignagomez2019} (Methods).\\
\\
However, the final supernova following the pulses can only meet the observed energy budget of SN2016aps if it forms a millisecond magnetar \cite{woosley2017}, which then accelerates the ejecta as it spins down. The massive pre-explosion core may require very rapid rotation to avoid collapse to a black hole, but a merger in a binary could conceivably spin up the star to facilitate this process. A millisecond magnetar could also directly contribute to (or even dominate) the radiative output through its spin-down luminosity, relaxing the requirement for a very massive ejecta, but the observed spectrum still requires massive CSM ejected shortly before explosion. Several hydrogen-poor SLSNe have shown evidence of CSM shells at larger radii \cite{yan2017,lunnan2018,chen2018}, indicating that engine formation is still possible following mass ejection. The need for a magnetar lowers the predicted rate of SN2016aps-like events, likely by an order of magnitude (Methods). The implied rate of $\sim10^{-4}$ per core-collapse SN and $\sim10^{-1}$ per SLSN IIn is roughly consistent with not having detected such an event until now.\\
\\
An exciting alternative is a `full' PISN interacting with a dense environment (a non-interacting PISN, even from a very massive star, cannot reach the observed luminosity \cite{sukhbold2016}). Blue supergiant progenitors with zero-age main sequence masses $150-175$\,\M\ can retain sufficiently massive helium cores ($64-84$\,\M) to encounter a terminal PISN explosion on their second pulse, following an initial failed PISN that expels only the hydrogen envelope (a more massive analogue of PPISNe) \cite{kasen2011,chatzopoulos2012b}. The kinetic energy of $\sim10^{52}$\,erg can power an extremely bright interaction. To retain a hydrogen envelope until explosion likely requires a merger in this case too, but avoiding excessive wind losses from the very massive post-merger remnant may necessitate merging only after core helium burning. With such fine-tuning, the predicted rate in this case is $\sim 2\times10^{-5}$ of the core-collapse rate \cite{vignagomez2019}.\\
\\
Detailed simulations will confirm whether SN2016aps is a PPISN, or even the less likely case of an interacting PISN. This event opens up new possibilities for finding very massive explosions at high redshift, being much brighter than either non-interacting PISNe or PPISNe without a central engine. The brightest radioactively-powered PISNe, from $130$\,\M\ helium cores, are faint at rest-frame UV wavelengths due to iron group absorption \cite{kasen2011}. The Large Synoptic Survey Telescope (LSST) will be able to detect a radioactive PISN at $z\lesssim0.75$, whereas a SN2016aps-like event could be detected out to redshift $z\gtrsim2$ (Methods, Extended Data Figure 4). This increases the volume over which these massive stars can be detected by a factor $\approx 7$. The PISN candidate SN2213-1745 at redshift $z=2.05$ \cite{cooke2012} may have been an analogue of SN2016aps, but was not observed spectroscopically. The upcoming \textit{James Webb Space Telescope} could spectroscopically classify a SN2016aps-like event at $z\gtrsim5$, offering a means to directly probe the deaths of first-generation stars.
\\
\\
\textbf{Acknowledgements}  M.N. is a Royal Astronomical Society Research Fellow. The Berger Time-Domain Group acknowledge NSF grant AST-1714498 and NASA grant NNX15AE50G. R.L. acknowledges a Marie Sk\l{}odowska-Curie Individual Fellowship within the Horizon 2020 European Union Framework (H2020-MSCA-IF-2017-794467). W.F. and K.P. acknowledge support from NSF Grant Nos. AST-1814782 and AST-1909358. The Margutti group acknowledges NSF Grant No. AST‐1909796, NASA grants 80NSSC19K0384 and 80NSSC19K0646. A.A.M. is supported by the LSST Corporation, the Brinson Foundation, the Moore Foundation via the LSSTC Data Science Fellowship Program, and the CIERA Fellowship Program. A.V-G. acknowledges support by the Danish National Research Foundation (DNRF132). Data were obtained via the NASA/ESA Hubble Space Telescope archive at the Space Telescope Science Institute, the Swift archive, the Smithsonian Astrophysical Observatory OIR Data Center, the MMT Observatory, the MDM Observatory, the Gemini Observatory, operated by the Association of Universities for Research in Astronomy, Inc., under agreement with the NSF, and the W.M.~Keck Observatory, operated as a partnership among the California Institute of Technology, the University of California, and NASA. Operation of the Pan-STARRS1 telescope is supported by NASA under Grants NNX12AR65G and NNX14AM74G. The authors respect the very significant cultural role of Mauna Kea within the indigenous Hawaiian community.
\\
\\
\textbf{Author contributions} M.N.~wrote the manuscript, led the analysis and obtained the HST and Gemini data. P.K.B.~devised the selection algorithm and identified SN2016aps as an interesting source, and analysed the HST images. E.B.~advised on the manuscript and leads the overall project. R.C.~and K.B.~obtained the MDM spectrum and classified SN2016aps. R.M.~analysed the UVOT data. P.K.B., S.G., A.B.~and P.C.~obtained FLWO data. R.L.~obtained the Keck spectra. A.V.-G.~did the rate calculations. A.A.M., F.J.M.~and R.R.L.~provided PTF data. W.F.~and P.K.B.~obtained MMT imaging. G.T., A.A.M.~and K.P.~obtained Keck imaging. All authors helped with the interpretation.
\\
\\
\textbf{Data availability} All data will be made publicly available upon publication, via the Open Supernova Catalog \cite{guillochon2017} and Weizmann Interactive Supernova Data Repository \cite{yaron2012}.
\\
\\
\textbf{Code Availability} MOSFiT \cite{guillochon2018} is publicly available at \url{https://github.com/guillochon/MOSFiT}. SuperBol \cite{nicholl2018} is publicly available at \url{https://github.com/mnicholl/superbol}.

\clearpage

\clearpage

\appendix
\renewcommand\thefigure{Extended~Data~\arabic{figure}}
\setcounter{figure}{0}

\begin{figure*}
    \centering
    \includegraphics[width=12cm]{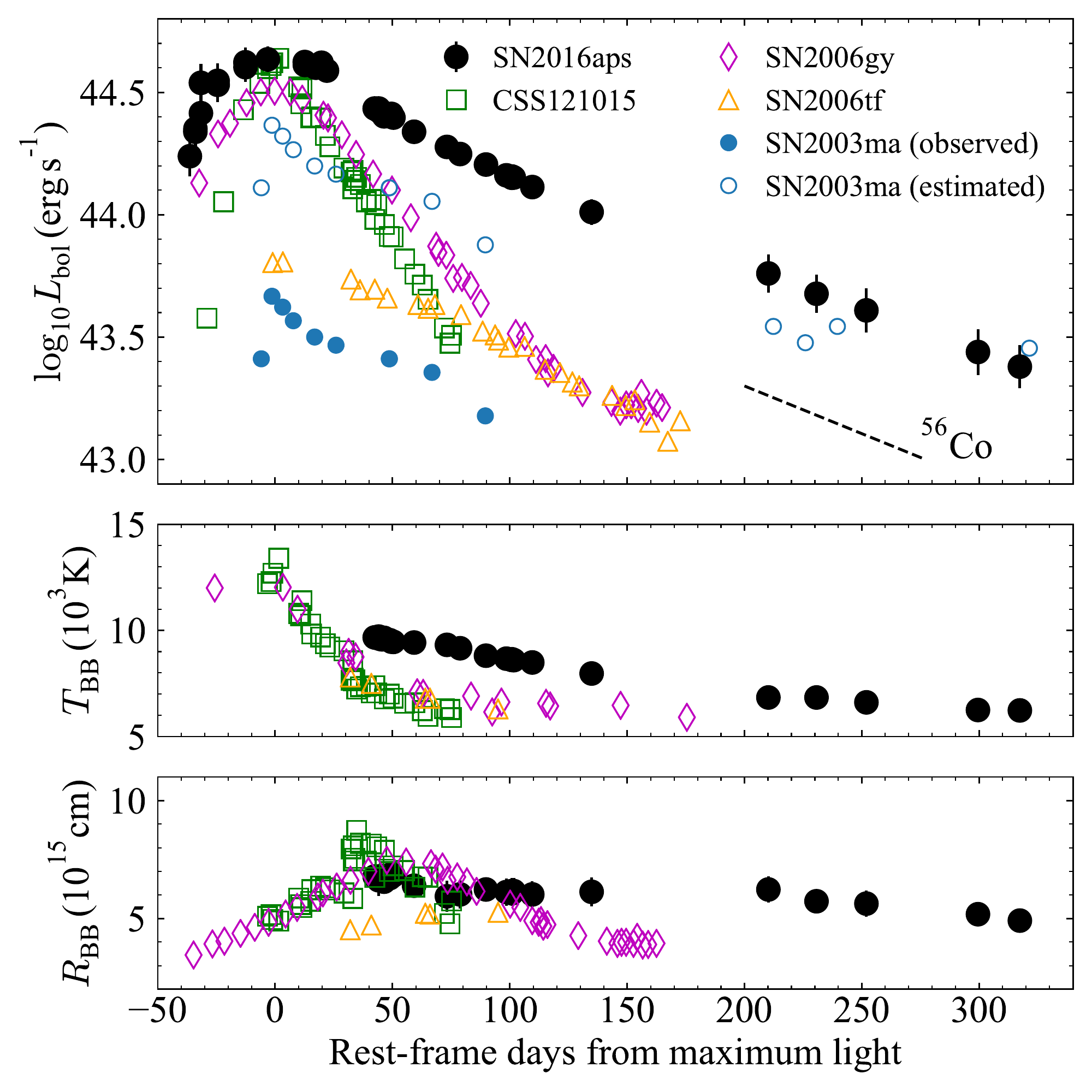}
    \caption{
    Bolometric light curve of SN2016aps.
    Top: Comparison of the bolometric light curve to other SLSNe IIn. The integrated luminosity is greater than any previously known SN. Middle: Temperature evolution from blackbody fits (Methods). Bottom: Photospheric radius from blackbody fits. Error bars show 1$\sigma$ uncertainties.
}
\end{figure*}

\clearpage

\begin{figure*}
    \centering
    \includegraphics[width=16cm]{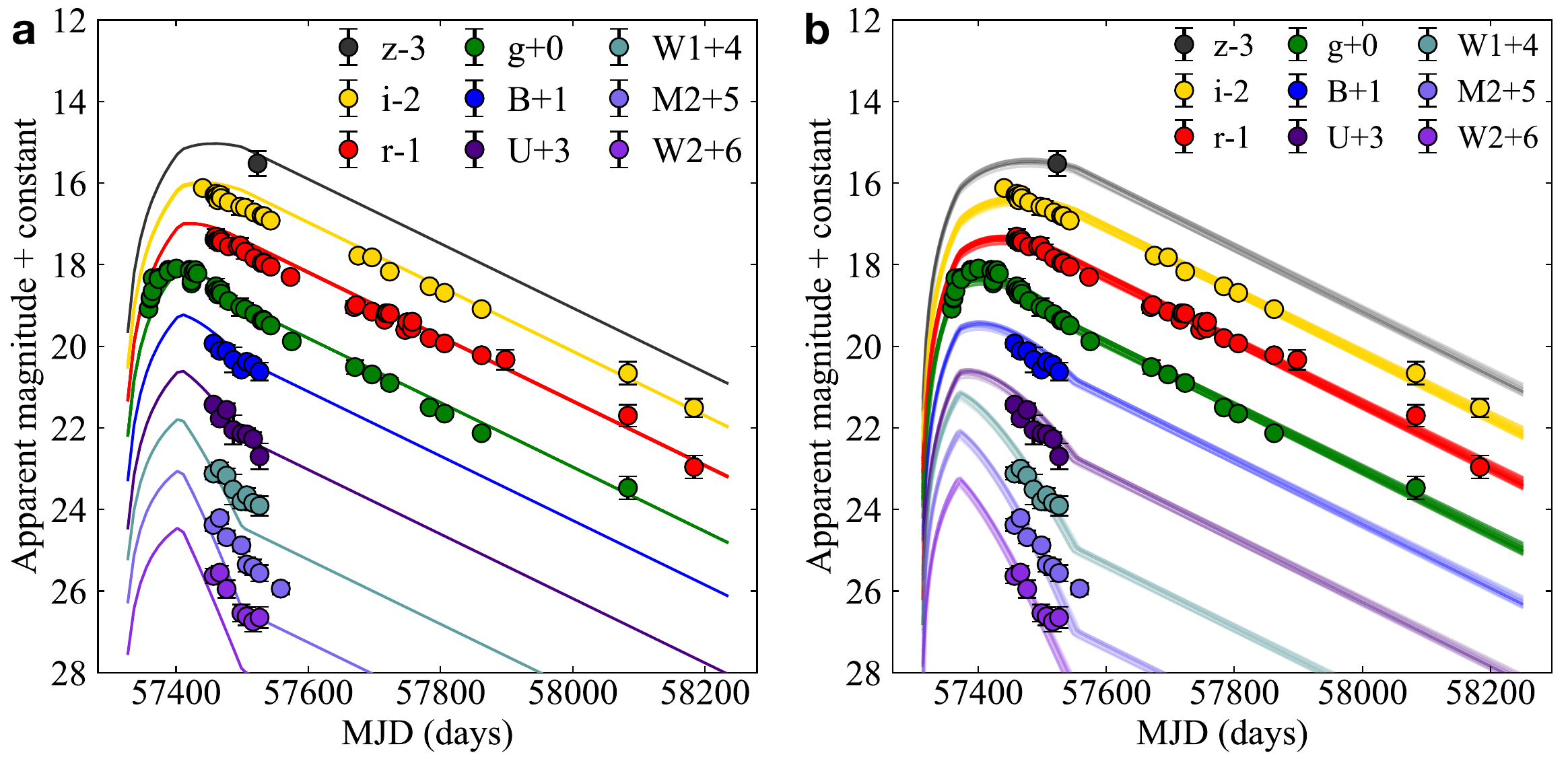}
    \caption{
    Fit to the light curve of SN2016aps with MOSFIT using a wind-like density profile. (a) For fixed parameters based on simple scaling relations (see Supplementary Information). The fit is reasonable overall but systematically under-predicts the UV bands. (b) Realizations of a full MCMC fit. This matches the UV but favours masses larger by a factor $\sim3$. Posteriors of the model parameters are given in Extended Data Figure 3. Error bars show 1$\sigma$ uncertainties.
}
\end{figure*}

\begin{figure*}
    \centering
    \includegraphics[width=16cm]{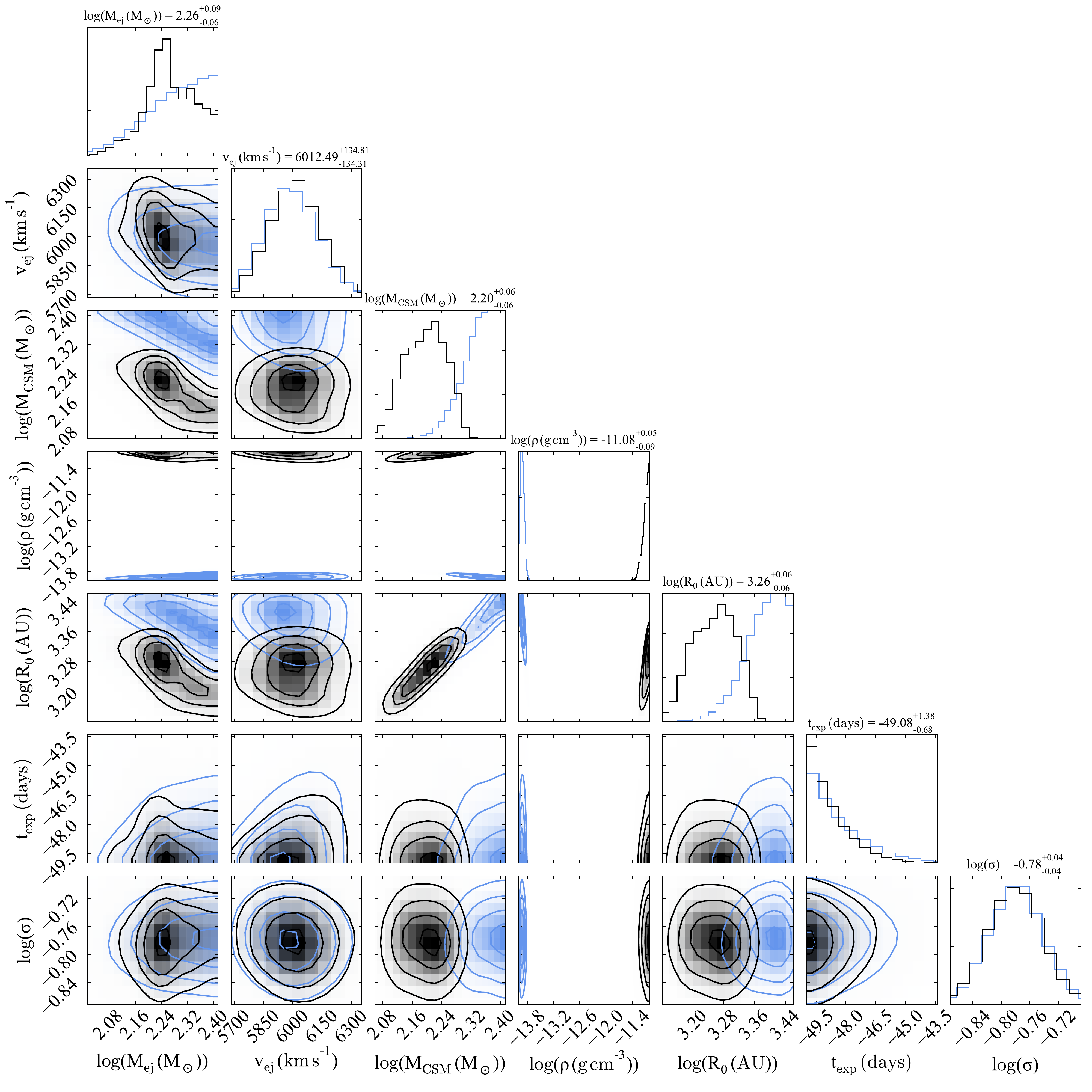}
    \caption{
    Posteriors for physical parameters inferred using the MOSFIT CSM model. Parameters are shown for fits using both shell ($\rho=$constant; black) and wind ($\rho\propto r^{-2}$; blue) density profiles. The additional variance (noise) parameter, $\sigma\sim0.1$, indicates a similar quality of fit independent of the assumed density.  Both models favour ejected masses $\gtrsim100$\M, with a comparable mass of CSM. Drawing from the joint $M_{\rm ej}-v_{\rm ej}$ posteriors gives a kinetic energy $E_k\approx5\times10^{52}$ erg in both cases.
}
\end{figure*}

\begin{figure*}
    \centering
    \includegraphics[width=16cm]{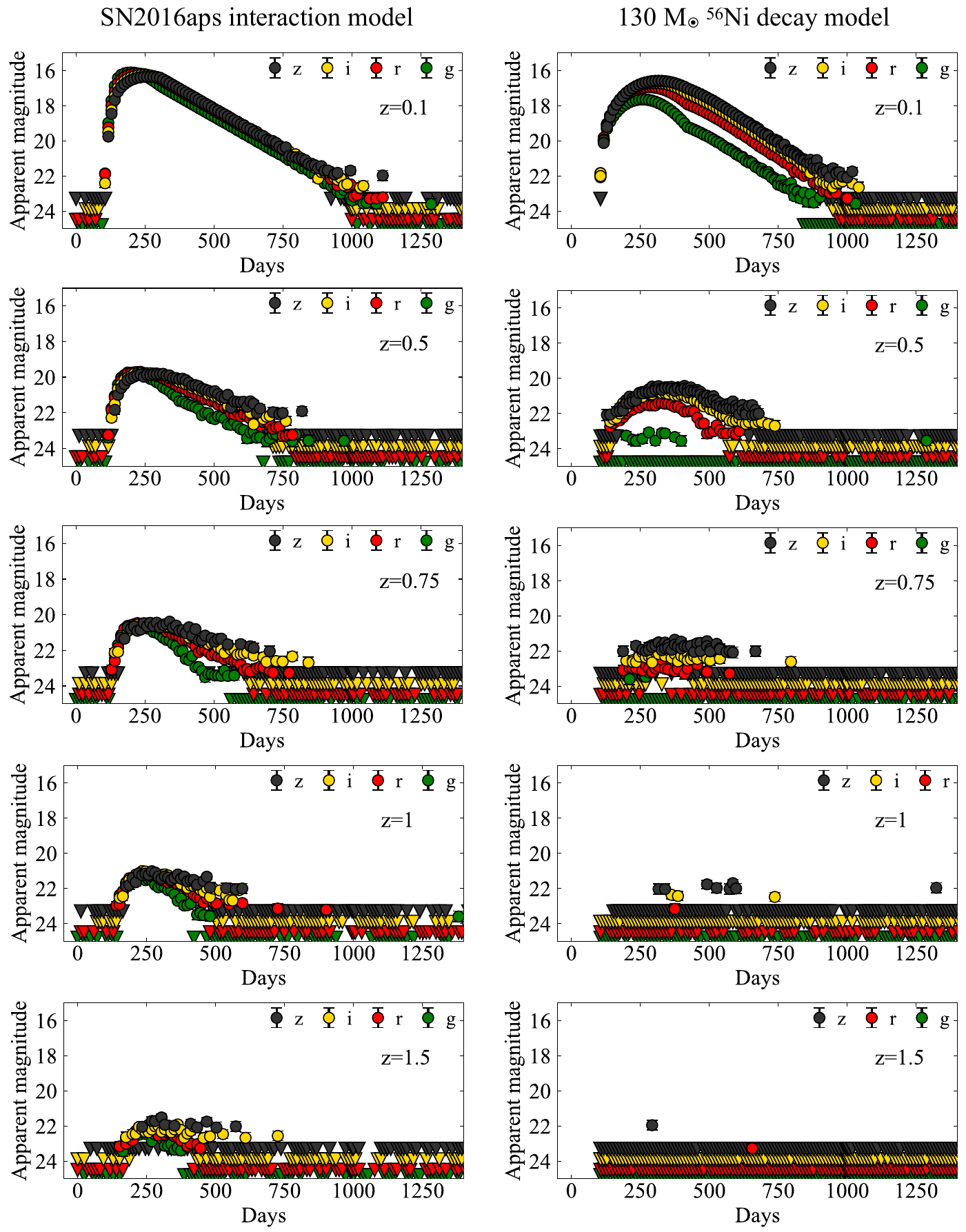}
    \caption{
    Simulated observer-frame LSST light curves in $g$,$r$,$i$,$z$ bands. The left column shows our interaction model for SN2016aps, while the right column shows a radioactively-powered PISN model for a 130\,\M helium core (Methods). The rows show the same models at redshifts $z=0.1$, 0.5, 0.75, 1 and 1.5. The interacting model is still well detected at $z=1.5$, because it is bright in the UV (whereas the radioactive model is heavily absorbed by metal lines). Therefore interacting events like SN2016aps are detectable over a volume that is larger by a factor $\gtrsim7$. Error bars show simulated 1$\sigma$ uncertainties.
}
\end{figure*}

\begin{figure*}
    \centering
    \includegraphics[width=12cm]{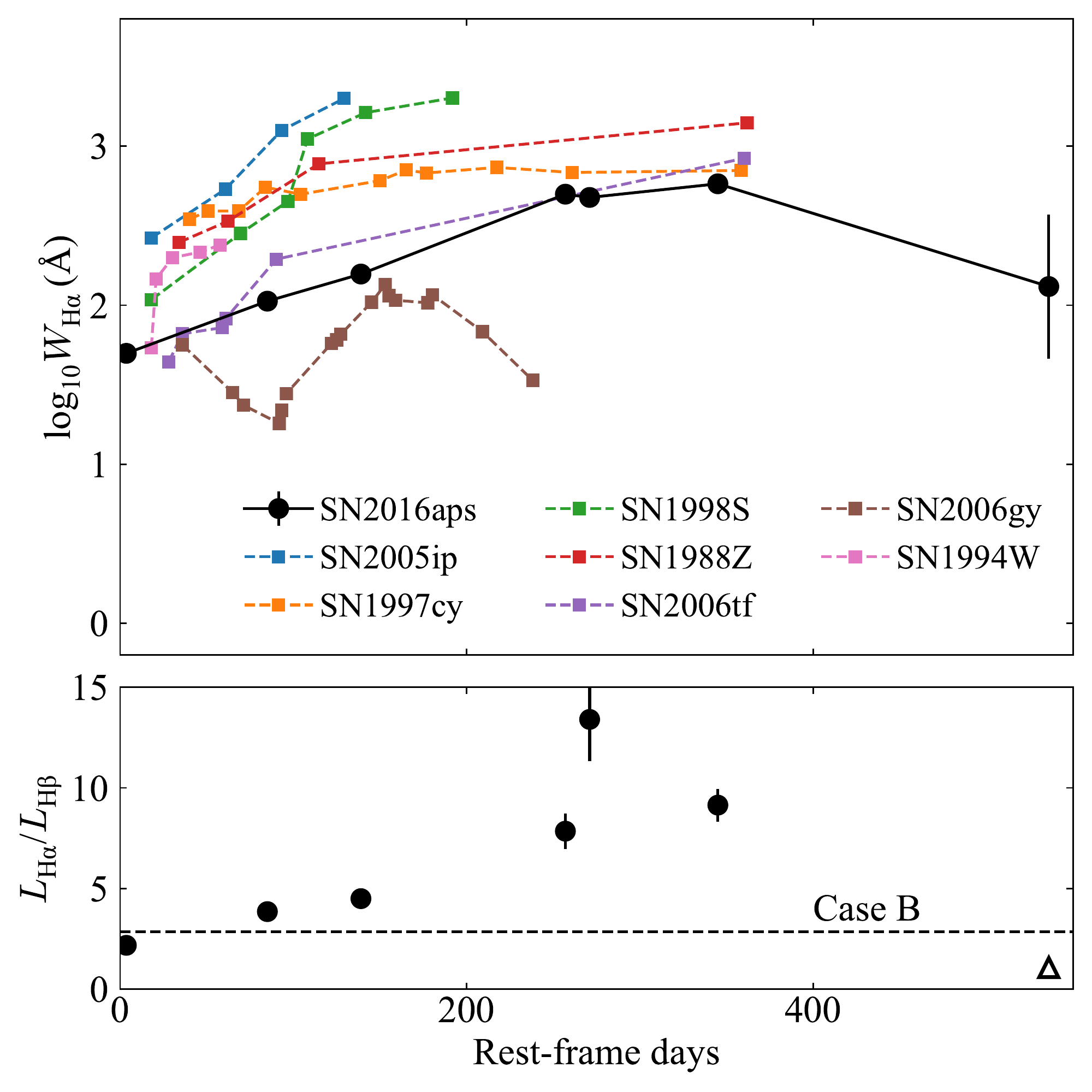}
    \caption{
    Measurements of the Balmer lines. The equivalent width of H$\alpha$ increases over the first 300 days as the continuum fades, similar to other Type IIn SNe and SLSNe, before decreasing in the final epoch. The H$\alpha$/H$\beta$ ratio is initially consistent with recombination (horizontal line), but at later times increases to $>10$, indicating collisional excitation. No H$\beta$ could be measured in the final spectrum, the arrow indicates a lower limit on this ratio. Error bars show 1$\sigma$ uncertainties.
}
\end{figure*}

\begin{figure*}
    \centering
    \includegraphics[width=12cm]{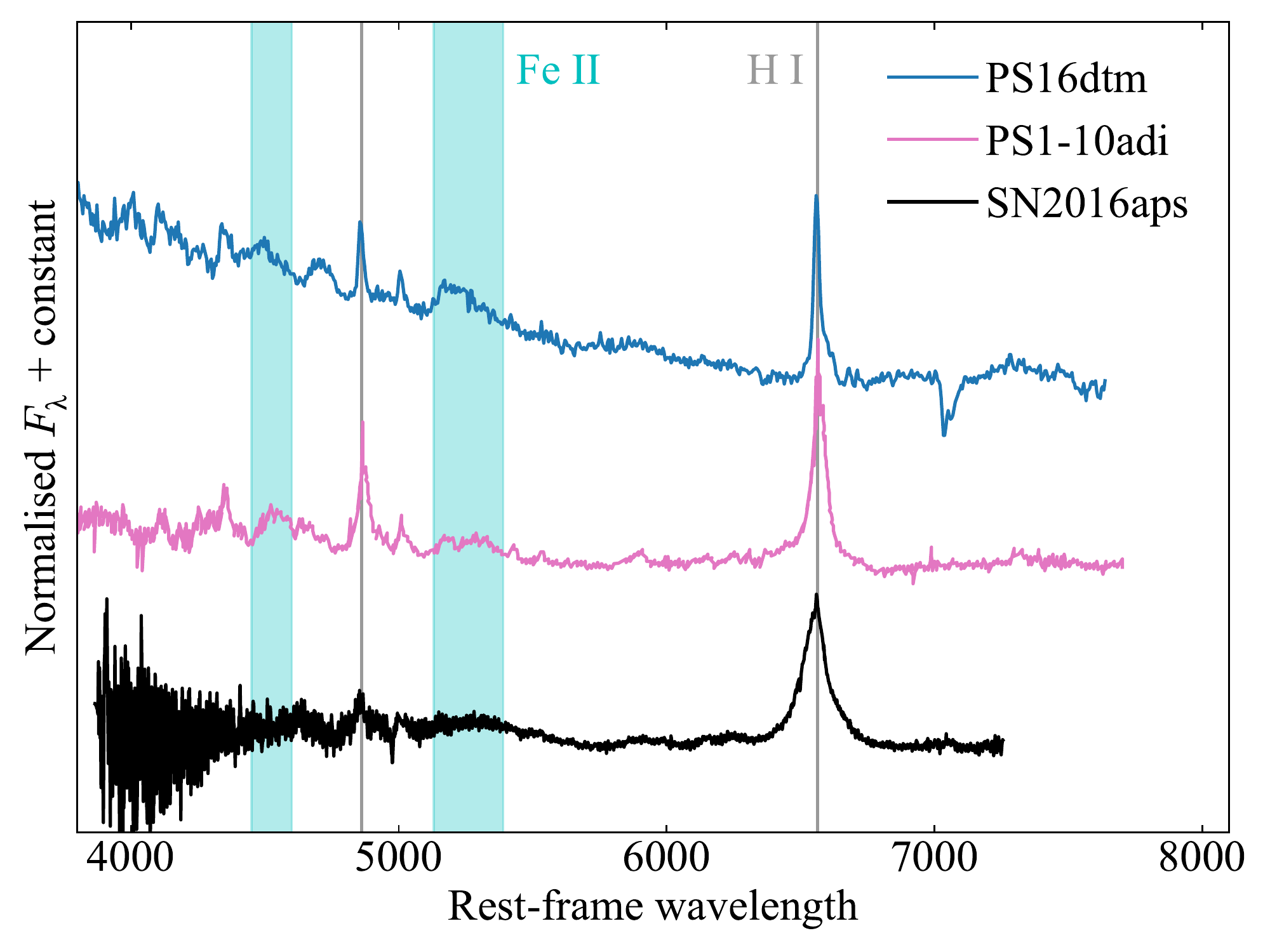}
    \caption{
    Spectroscopic comparison of SN2016aps to the energetic nuclear transients PS16dtm and PS1-10adi. PS16dtm is thought to be a TDE \cite{blanchard2017} and PS1-10adi has been suggested to be a possible SN close to an AGN \cite{kankare2017}, these two nuclear transients closely resemble each other. The spectra shown here are at around 200 days after maximum light. SN2016aps is distinguished from these events by broader and more symmetrical Balmer lines (lacking a red shoulder), and a lack of strong, narrow Fe II emission. SN2016aps also lacks the [O III] emission seen at 5000\,\AA\ (see also Supplementary Information). The apparent absorption in PS16dtm at 7000\,\AA\ is a telluric feature from the Earth’s atmosphere.
}
\end{figure*}

\clearpage

\renewcommand\thefigure{S\arabic{figure}}
\setcounter{secnumdepth}{0}

\section{Methods}

\textbf{Spectroscopy}
We observed SN2016aps spectroscopically using the Ohio State Multi-Object Spectrograph (OSMOS) \cite{martini2011} on the 2.4-m Hiltner telescope at MDM observatory, the FAST spectrograph on the FLWO 1.5-m telescope \cite{fabricant1998}, the Blue Channel spectrograph on MMT \cite{schmidt1989}, GMOS \cite{hook2004} on Gemini North, and the Low Resolution Imaging Spectrograph (LRIS) \cite{Oke1995} on the Keck I 10-m telescope.  The majority of these data were reduced in \textsc{Pyraf}, applying bias subtraction, flat-fielding, and extraction of the 1-dimensional spectrum. Wavelength solutions were derived using arc lamps, and flux calibration achieved using observations of standard stars. Keck spectra were reduced using the dedicated \textsc{lpipe} package \cite{perley2019}.
We corrected all spectra for a foreground extinction of $E(B-V)=0.0263$ \cite{schlafly2011} and for a cosmological redshift $z=0.2657$. We assumed a Planck cosmology in all distance calculations \cite{Planck2016}.

Fits to the line profiles with Gaussian and Lorentzian functions were conducted using \textsc{scipy}. We approximated the local continuum by fitting a linear function to the following relatively line-free regions. For H$\alpha$, we used 6100-6300\AA\ and 6700-6800\AA\, and for H$\beta$ we used 4600-4780\AA\ and 4980-5000\AA. The profiles are well fit with a Lorentzian function, indicating that broadening is due to electron scattering rather than expansion. Fluxes and equivalent widths were obtained by direct integration. The flux ratio between the H$\alpha$ and H$\beta$ lines evolves from $2-3$ during the first 50 days, to $\gtrsim 7-10$ after 200 days (Extended Data Figure 5). The early ratio is consistent with hydrogen recombination in ionised CSM, while the late emission can be excited \cite{smith2010} by the passage of shock fronts through the CSM and ejecta. SN2006gy initially exhibited a similar evolution, with an early ratio of $\approx3$, but at times $\gtrsim100$ days the ratio increased to more extreme values $\sim100$. However, a direct comparison of these values with SN2016aps may be misleading, as SN2006gy showed significant absorption components (both narrow and broad) in both emission lines.

The spectra shown in Figure \ref{fig:spec} have been smoothed using a Savitsky-Golay filter \cite{savitsky1964}. They are initially dominated by a blue continuum superposed with hydrogen Balmer emission lines, typical \cite{smith2007a,benetti2014} of these events. As the spectra evolve, SN2016aps most closely resembles \cite{smith2008,fransson2014} long-duration SLSNe IIn such as SN2010jl and SN2006tf. SLSNe IIn also share some spectroscopic similarities with energetic nuclear transients \cite{blanchard2017,kankare2017}, however we argue in the Supplementary Information that these are physically distinct classes. We show a spectroscopic comparison in Extended Data Figure 6, highlighting significant differences in the Balmer profiles and iron lines.
\\
\\
\textbf{Photometry}
Optical photometric observations of SN2016aps in $g,r,i$ bands were obtained using KeplerCam on the 1.2-m telescope at Fred Lawrence Whipple Observatory (FLWO), MMTCam on the 6.5-m MMT telescope , the Gemini Multi-Object Spectrograph (GMOS) on the 8-m Gemini North telescope, and DEIMOS on the 10-m Keck II telescope. Images were reduced using \textsc{pyraf} to apply bias subtraction and flat-fielding. Dark correction was also performed for the MMTCam images. Photometry was measured with a custom wrapper for \textsc{daophot}, using nearby stars from PanSTARRS Data Release \cite{Flewelling2016} 1 to determine the point-spread function (PSF) and photometric zeropoint of each image. At later epochs, FLWO images from neighbouring nights were co-added to improve the signal-to-noise ratio.

We subtracted the underlying host galaxy flux from each measurement using the galaxy magnitudes measured in late-time imaging. In the $g$ band, where no host imaging is available, we assumed $g_{\rm host} \sim 24.5$\,mag by interpolating across the $u,r$ and $i$ bands. We assume a 20\% uncertainty on the host flux in all cases. The change in magnitude following host subtraction becomes significant ($>0.1$\,mag) only after $\sim 400$ days of the light curve fading, and therefore has no effect on our estimate of the total luminosity from SN2016aps.

Publicly available Pan-STARRS data in $i$ and $z$ bands, obtained as part of the Pan-STARRS Survey for Transients \cite{huber2015}, were downloaded from the Pan-STARRS Transient Science Server hosted at Queen's University Belfast. These magnitudes are measured by the Pan-STARRS Data Processing System \cite{magnier2016} after subtraction of a reference image, and hence are free of host galaxy light.

Additional optical photometry was obtained with the CFH12K camera \cite{Rahmer08} on the Palomar Observatory 48-inch telescope \cite{law09} (P48). Images were processed by the IPAC image-subtraction pipeline, which subtracts background galaxy light using deep pre-SN images and performs forced point-spread function (PSF) photometry at the location of the SN  \cite{Masci17}. The photometry is then calibrated to the PTF photometric catalog \cite{Ofek12}. We estimate the time of maximum light as MJD 57404 (2016 January 17 UT) using a 4th-order polynomial fit to the PTF $g$-band data.

We also carried out imaging using the Neil Gehrels \textit{Swift} Observatory with the UV-Optical Telescope (UVOT) in the \textit{UVW2}, \textit{UVM2}, \textit{UVW1}, $U$, $B$, and $V$ filters. We extracted the SN flux in each image using a 3" aperture and following the recommended procedures \cite{Brown2009}, and calibrated to Vega magnitudes in the UVOT photometric system \cite{Breeveld2011}. As UV imaging covers only the first $\sim 100$ days, we assume the host contribution to be negligible in these observations.

When analysing the light curves we accounted for a Galactic extinction $E(B-V)=0.0263$ \cite{schlafly2011}, but assumed internal extinction within the host galaxy was negligible. The galaxy appears to be a dwarf galaxy similar to the hosts of SLSNe I, which generally have low extinction (consistent with zero in many cases) \cite{lunnan2014}. At early times, the Balmer decrement (H$\alpha$/H$\beta$) in the spectra of SN2016aps is consistent with the expected value for Case B recombination, which lends support to a low internal extinction. Finally, we do not see a strong Na I absorption line (thought to be correlated with dust extinction) \cite{poznanski2012} from the host galaxy. Including a significant host extinction would serve to increase the total luminosity of SN2016aps even further, so would only strengthen the results presented here.

To obtain the absolute $r$ and $u$-band light curves in Figure \ref{fig:lc}, we used the \textsc{s3} package \cite{inserra2018} to derive cross $K$-corrections from our spectra. We linearly interpolated these corrections to epochs with photometry. At this redshift, observed $i$-band was closest to rest-frame $r$, and observed $g$ to rest-frame $u$. The comparison data \cite{rest2011,benetti2014,fransson2014,smith2008,ofek2007,agnoletto2009,smith2010,miller2010,moriya2019,tartaglia2019,arcavi2017,terreran2017} in the figure were obtained from the Open Supernova Catalog \cite{guillochon2017} if possible, or otherwise directly from these papers.

The bolometric light curve was calculated using \textsc{superbol} \cite{nicholl2018}, including extinction corrections, interpolation to a common set of epochs, transformation to the rest-frame and blackbody fits. The bolometric light curve, (on log-linear scale), the derived temperature and radius evolution, and comparisons to other events, are shown in Extended Data Figure 1. We assume constant colours prior to discovery, as we only have PTF $g$-band data at early times. If the photospheric temperature was higher during the rising phase, as is often the case in SNe, the total luminosity would be even greater. We also note that we do not have near-infrared data to look for dust formation, common in interacting SNe at late times \cite{gall2014,bhirombhakdi2019}. Any late-time infrared excess, as seen \cite{gall2014,fransson2014,tartaglia2019} in the spectroscopically similar and slowly-evolving SLSNe IIn SN2010jl and SN2015da, would also increase the total $E_{\rm rad}$ further. Thus the integrated observed $E_{\rm rad}=5\times10^{51}$\,erg is a conservative lower limit on the total energy.

To power the peak luminosity with radioactive decay would require $\simeq 20$\,\M\ of $^{56}$Ni. This decays to $^{56}$Co on a half-life of 6 days then to $^{56}$Fe on a half-life of 77 days; thus at peak the energy would be primarily from $^{56}$Co decays. Although we favour circumstellar interaction as the power source in SN2016aps, we note that a radioactively powered light curve still requires a huge progenitor mass consistent with a PISN: to produce 20\,\M\ of  $^{56}$Ni needs a core mass \cite{kasen2011} $\gtrsim120$\,\M.
\\
\\
\textbf{X-rays}
We imaged SN2016aps with the X-ray Telescope (XRT) on-board \textit{Swift}. Stacking all the data for a total exposure time of 37\,ks obtained between MJD 57456.7- 57558.4, we detect no X-rays to a limiting count-rate of $<4.2\times10^{-4}$\,ct\,s$^{-1}$. We use the online WebPIMMS tool to convert this to flux, assuming a thermal Brehmsstrahlung spectrum with $kT=20$\,keV, similar to SN2014C \cite{margutti2017b} (one of the best-observed interacting SNe at X-ray wavelengths), and a Galactic hydrogen column density of $2.27\times10^{20}$\,cm$^{-2}$ in the direction of SN2016aps. The unabsorbed flux is $F_X<2.0\times10^{-14}$\,\ergs\,cm$^{-2}$, corresponding to a luminosity $L_X<4.7\times10^{42}$\,\ergs\ (0.3-10\,keV). Taking our peak bolometric luminosity from the UV-optical data, this implies $L_X / L_{\rm bol} < 0.011$. SN2010jl had $L_X / L_{\rm bol} \approx 0.01$, argued to be low due to attenuation of X-rays from the shock by the optically thick CSM \cite{fransson2014}. Our measurement therefore implies that SN2016aps exhibits at least as much X-ray attenuation as SN2010jl.
\\
\\
\textbf{HST imaging and host galaxy properties}
We obtained late-time data using the \textit{Hubble Space Telescope} (\textit{HST}) on 2019-07-27 UT, corresponding to 1017 days after maximum light in the SN rest-frame (Program ID: 15709, PI: Nicholl). Drizzled data were downloaded from the Mikulski Archive for Space Telescopes. We used the F775W filter on the Advanced Camera for Surveys (ACS) and the F390W filter on Wide Field Camera 3 (WFC3). We matched the F390W image, and an earlier image from MMTCam, to the F775W image using 12 common sources. The uncertainty in the astrometric tie is 0.0282" between the ground- and space-based images, and 0.0061" between the two \textit{HST} images. We determine the SN position to be 10h19m02.124s, +74$^\circ$42'24".82 in the system of the F775W image, using \textsc{sextractor}, with an uncertainty of 0.0084".

The host galaxy, previously undetected in ground-based surveys, is easily identified in the \textit{HST} images. We measure integrated host galaxy AB magnitudes of $m_{F775W} = 23.7\pm0.09$\,mag and $m_{F390W} = 24.9\pm0.07$\,mag, where we have checked using PanSTARRS DR1 sources that any deviations of the \textit{HST} imaging zeropoints are smaller than the photometric errors.

An unresolved source at the same position in Keck images obtained on 2019-02-26 has an $i$-band measurement consistent with the ACS magnitude, we therefore assume that this measurement is host-dominated. An $r$-band image on the same night gives $m_r=23.9\pm0.3$\,mag.

We used \textsc{galfit} \cite{peng2002} to measure the physical size of the galaxy, finding an effective radius of $R_e = 2.1$\,kpc with an axis ratio $b/a=0.27$ in the F775W image. We also find that SN2016aps is offset from the center, as measured in red/optical light, by 3.1 \textit{HST} pixels, or 0.15". This is greater than the uncertainty in the astrometric tie between the SN and host images. Thus SN2016aps is inconsistent at the $\approx 5\sigma$ level with having occurred in the nucleus of its host galaxy.

The F775W filter is very close to rest-frame $r$ band at this redshift. For an inferred absolute magnitude $M_r\simeq-17$\,mag, we estimate the galaxy stellar mass as $M_*/M_\odot \sim L/L_\odot$, where $L=10^{0.4(M_{\odot,r}-M_r)}$, giving $M_*\sim 10^{8.6}$\,\M. Using the mass-metallicity relation from \cite{andrews2013}, this implies a metallicity $Z\sim0.4 Z_\odot$. A more accurate determination of the host galaxy properties will require deep spectroscopy and imaging over a broad wavelength range after we are more confident that SN2016aps has completely faded.

The UV luminosity and spatial extent of the host indicate a mean star-formation rate surface density of $\approx 0.04$\,\M\,yr$^{-1}$ \cite{kennicut1998}, consistent with the lower end of the distribution measured for Type I (hydrogen-poor) SLSNe \cite{lunnan2015}. Those explosions are known to favour the brightest UV (most star-forming) regions of their hosts, suggesting young, massive progenitors \cite{lunnan2015}. The association of SN2016aps with the UV-brightest region of its host thus points to a massive star progenitor.
\\
\\
\textbf{Rate estimates for interacting (P)PISNe}
For each of our suggested progenitor channels, we estimate the rates of forming core masses in the necessary range via binary mergers using a pre-established method \cite{vignagomez2019}. The rate is given by:
\begin{equation}
    R = f_{\rm bin} \times f_{1} \times f_{2} \times f_{\rm sep},
\end{equation}{}
where $f_{\rm bin}\simeq0.7$ is the fraction of massive stars in close binaries \cite{sana2012}. The next factor $f_{1}$ is the fraction of primary stars sufficiently massive to form the desired core mass post-merger (modelled using Modules for Experiments in Stellar Astrophysics \cite{paxton2011,paxton2018}), normalised to the core-collapse SN rate assuming a Salpeter initial mass function and that stars with masses in the range $8-40$\,\M\ experience CCSNe. The factor $f_{2}\approx0.15$ is the fraction of secondary stars, given a suitable primary, that are sufficiently massive for this channel \cite{vignagomez2019}, and $f_{\rm sep}$ is the fraction of suitable binaries with the appropriate separation to merge during a given evolutionary phase. The largest uncertainty is on $f_{2}$, which can be lower by an order of magnitude for more pessimistic assumptions requiring near-equal-mass binaries \cite{vignagomez2019}.

For the PPISN channel, we look for existing models \cite{woosley2017} that begin pulsing between $\sim {\rm few} \times 0.1-10$ years before core collapse. This corresponds to a range in helium core mass of $\simeq 40-50$\,\M. In this case we find (from the MESA models) $f_1=0.03$. We assume the merger can happen at any point after core hydrogen burning; for a log-flat distribution of separations and a maximum stellar radius of $\approx 760\,{\rm R}_\odot$, this gives $f_{\rm sep}=0.44$. Thus the overall rate of formation of suitable progenitors is $R_{\rm PPISN}=0.7\times0.03\times0.15\times0.44=1.3\times10^{-3}$ per CCSN. The rate of SN2016aps-like transients will include another factor accounting for those progenitors with sufficiently rapid core rotation to form magnetars. We estimate $f_{\rm mag}\sim0.1$, i.e.~the fastest 10\% of rapidly rotating massive stars can produce magnetars \cite{yoon2006,demink2013,nicholl2017}, giving an overall rate $R_{\rm PPI+mag}\sim10^{-4}$ per CCSN.

For the more massive PISN channel, the core mass range of interest \cite{kasen2011} is $\approx 64-84$\,\M, which gives $f_1=0.02$. However, in this case it is less clear whether the massive merger product can retain its hydrogen envelope unless the merger happens late (after core helium burning), giving a much narrower range of allowed separations ($f_{\rm sep}\sim0.01$). In this case, the estimated rate is $R_{\rm PISN+CSM}=2.1\times10^{-5}$ per CCSN. Thus the PPISN channel appears the more likely, even with its requirement for a central engine.
\\
\\
\textbf{Detectability with LSST and JWST}
We use our light curve model (see also Supplementary Information) to estimate the observability of a transient like SN2016aps for next-generation instruments. \textsc{mosfit} provides a simple built-in routine to estimate the signal-to-noise for a transient observation given a specified filter and limiting magnitude. We generate synthetic light curves in the $g,r,i$ and $z$ bands taking the limiting magnitudes appropriate for the Large Synoptic Survey Telescope (LSST) \cite{lsst2009}: $g_{\rm lim}=24.8$\,mag, $r_{\rm lim}=24.5$\,mag, $i_{\rm lim}=23.9$\,mag and $z_{\rm lim}=23.3$\,mag. Observations in the $u$ and $y$ bands are shallower so we do not consider them here. We place the best-fitting CSM shell model at redshifts $z=0.1,0.5,0.75,1,1.5$ for these simulations.

We perform the same calculation for a massive radioactively-powered PISN model, based on a 130\,\M\ helium core. Our \textsc{mosfit} implementation uses an ejecta mass of 130\,\M, a radioactive nickel mass of 39\,\M, and an ejecta velocity of 8000\,\kms. Absorption by heavy elements is implemented via a linear cut-off \cite{nicholl2017} in the blackbody spectral energy distribution below 5000\,\AA, to mimic the red spectra from more detailed simulations \cite{kasen2011,dessart2012,jerkstrand2016}.

The results are shown in Extended Data Figure 4. While both models are detectable at low-redshift for 2-3 years in the LSST survey, the situation is very different at higher redshift. Radioactive PISNe are only detected at $z\lesssim0.75$ as above this the bulk of their emission is redshifted out of the optical bands. However a SN2016aps-like massive interacting event can be detected as far as $z\sim2$, as they do not suffer from the same rest-frame UV absorption (the power source is not coupled to heavy element production). Thus strong CSM can increase the volume over which PISNe are detectable by up to a factor $\sim7$, increasing our chances of finding PISNe with LSST. However, the distribution of CSM and ejecta properties is currently unknown, and the volumetric rate of PISNe is highly uncertain (but constrained \cite{cooke2012,nicholl2013} to be $<10-100$\,Gpc$^{-3}$\,yr$^{-1}$), making a quantitative estimate of the number of interacting PISNe unfeasible at this time.

We note that characteristics of a SN2016aps-like event at $z=2$ are consistent with the transient SN2213-1745, discovered \cite{cooke2012} in stacked Canada-France-Hawaii Telescope Legacy Survey data and confirmed to be in a galaxy at $z=2.05$. This event is one of the best candidates for a PISN due to its luminosity and slow light curve evolution, but it was found \cite{cooke2012} that the observed flux was brighter and bluer than massive $^{56}$Ni decay-powered PISN models \cite{kasen2011}. Interaction with a massive CSM can explain this blue flux excess, while the required mass is likely still consistent with a PISN. The peak apparent magnitude $r\approx24$ confirms that events like SN2213-1745 and SN2016aps will be detectable at $z\approx2$ with LSST.

We also calculate the observability of a SN2016aps-like transient with the \textit{James Webb Space Telescope} (\textit{JWST}). At $z=5$, SN2016aps would have reached $\approx24$\,mag in the NIRCam F410M filter (40,900\,\AA\ in the observer frame, corresponding to $\approx6,800$\,\AA\ in the rest frame). This matches the limiting magnitude for spectroscopy with NIRSpec to achieve a S/N ratio of 10 with the G495M disperser (covering $\approx 30,000-50,000$\,\AA). Thus we could comfortably detect the strong H$\alpha$ emission and spectroscopically classify a SN2016aps-like event with high confidence as far as $z\gtrsim5$.

\clearpage

\section{Supplementary Information}

\subsection{Ruling out a tidal disruption event}

One issue that arises in understanding the origin of the most luminous transients is the difficulty in distinguishing SLSNe from extremely bright tidal disruption events (TDEs), i.e. transients powered by the destruction of a star passing within the tidal radius of a supermassive black hole. ASASSN-15lh was previously thought to be the most luminous known SN \cite{dong2016}, but subsequent studies argued that its properties were more consistent with a TDE interpretation \cite{leloudas2016,margutti2017}.

In the case of SN2016aps, we can clearly rule out a TDE explanation. Most TDEs show a constant or even increasing temperature over time, whereas SN2016aps shows a decreasing temperature typical of SN cooling (Extended Data Figure 1). The spectrum is also typical of SLSNe IIn, in particular Lorentzian line profiles (Figure 2) and the evolution of the H$\alpha$ equivalent width with time (Extended Data Figure 5). SN2016aps is inconsistent at the $\approx5\sigma$ level with having occurred in the centre of its host, whereas a TDE would have occurred in the nucleus.

\subsection{Comparison to SLSNe IIn}
Figures 2 and 3 show spectroscopic and photometric comparisons with a number of SLSNe IIn. Although SN2016aps is more than twice as energetic as any of the other events, there are a number of similar properties among the class. Considering only events that emit $\gtrsim 10^{51}$\,erg (SN2006gy, SN2003ma, SN2008am, SN2008fz, SN2015da, CSS121015) or fade on a similar timescale to SN2016aps (SN2006tf, SN2010jl), all events show similar maximum light spectra, with blue continua and dominated by roughly symmetric, scattering-broadened (i.e.~Lorentzian) Balmer lines. This similarity persists for most events \cite{rest2011,chatzopoulos2011,drake2011,smith2008,tartaglia2019} throughout their evolution, as the continuum cools and the equivalent widths of the lines increase. SN2006gy and CSS121015 are exceptions. From around 90 days after explosion, SN2006gy displays a rather asymmetric H$\alpha$ line, with a serious of narrow P Cygni absorption lines from slow, unshocked CSM \cite{smith2010}, whereas CSS121015 instead shows broad metal lines resembling \cite{benetti2014} SLSNe I. These events may require a more complex CSM structure. However, the key point is that SN2016aps is not unusual spectroscopically for a SLSN IIn.

The host galaxy environments provide another point of overlap between many of these events. CSS121015, SN2008fz, SN2010jl and SN2006tf all occurred in dwarf galaxies, with absolute magnitudes $M_r\approx-17$ to $-18$ mag, comparable to SN2016aps. The host metallicities estimated from emission line diagnostics are $\lesssim 0.3-0.4 Z_\odot$ for SN2010jl \cite{stoll2011} and SN2008am \cite{chatzopoulos2011}. In contrast, SN2006gy, SN2003ma, and SN2015da occurred in more massive galaxies with $M\sim -21$ mag. SN2006gy exploded close to the center of a galaxy hosting an active galactic nucleus, whereas the others were significantly offset from the centers of their hosts \cite{rest2011,tartaglia2019}. There was no evidence of AGN variability in the host of SN2003ma in 7 years of pre-SN observations \cite{rest2011}. This will be important in the next section, when we compare to a population of transients that seem to occur exclusively in active galactic nuclei.

To summarise: of the eight SLSNe IIn that come closest to SN2016aps in photometric properties, all show similar maximum light spectra, with six showing similar spectra throughout the photospheric phases (nebular spectra are not available). Five events occurred in metal-poor or dwarf galaxy environments. Four events satisfied both of these similarity criteria. The spectroscopic consistency is not surprising, as interaction with an opaque, hydrogen-rich circumstellar shell can lead to similar spectra across a wide range of explosion parameters \cite{dessart2015}. Pulsational PISN mass ejections have been suggested as a plausible means to build up a massive CSM in other events \cite{smith2010}. \textit{The key difference with SN2016aps is that the total energy and long timescale demand an extremely high final explosion energy beyond what is possible with conventional core-collapse}. Thus while these events certainly form an observational class, and possibly a physical class if the CSM is produced by the same mechanism, SN2016aps requires an extra ingredient: either a significantly larger mass or an extremely energetic explosion.

Simulations of very energetic core collapse SNe leading to SLSNe exist \cite{umeda2008,moriya2010}, however the explosion energy is often a free parameter in these models. One physically-motivated way to get a large explosion energy is through rotation -- either in the form of a magnetar central engine accelerating the ejecta on the spin-down timescale \cite{kasen2010}, or the launching of jets in a collapsar-like model \cite{wheeler2000}. In the latter case at least, the explosion may be highly asymmetric \cite{couch2009}. Unfortunately, the presence of dense CSM in events such as SN2016aps obscures the geometry of the underlying ejecta.

\subsection{Comparison to hydrogen-rich transients in active galactic nuclei}
Recently a population of extremely energetic hydrogen-rich transients have been discovered in the centres of active galaxies \cite{drake2011,blanchard2017,kankare2017}. In particular, PS1-10adi radiated \cite{kankare2017} $\sim 2\times10^{52}$\,erg. They have been interpreted by some authors \cite{kankare2017,drake2011} as most likely resulting from SNe interacting with dense material in the centres of these galaxies, though they do not rule out TDEs as an alternative explanation. In this section we argue that the latter interpretation is more likely, and that despite superficial spectral similarity, the off-nuclear SN2016aps is distinct from these events.

Extended Data Figure 6 shows the spectrum \cite{kankare2017} of PS1-10adi at around 200 days after maximum, compared to our Gemini spectrum of SN2016aps at a similar phase. Although the strongest features in both spectra are the Balmer lines, these lines show a red shoulder in PS1-10adi but are symmetric (and slightly blueshifted) in SN2016aps, more typical \cite{fransson2014} of SLSNe IIn with electron scattering in an expanding atmosphere. The line profiles, and the presence of narrow Fe II emission, are much more similar to PS16dtm \cite{blanchard2017}. This is another transient in an AGN, but in this case there is strong evidence that the source is associated with the supermassive black hole, rather than being a SN.

In particular, historical observations from \textit{XMM-Newton} showed that the AGN is an X-ray source, but observations with \textit{Swift} during the optical flare showed that X-rays had faded by at least an order of magnitude. A SN is unable to obscure the AGN accretion disk, but formation of an atmosphere or disruption of the existing disk by a TDE naturally explains the X-ray fading. Interestingly, PS1-10adi showed X-rays that appeared 5 years after the optical flare, which could be due to the formation of a disk following the TDE, or an existing disk that is revealed after the debris settles down (the inverse of the process that explains PS16dtm).

Another discriminant between SN2016aps and this population of nuclear transients is the lack of a mid-IR echo. All of these event studied to date \cite{kankare2017}, including PS16dtm \cite{blanchard2017}, have shown very bright mir-IR emission, detected by the \textit{Wide-field Infrared Survery Explorer (WISE)}, following the optical flare \cite{jiang2019}. This has been interpreted \cite{jiang2019} as a TDE signature, as the mid-IR luminosity in these events is challenging to produce with a SN, but consistent with a dusty AGN torus. At the distance of SN2016aps, a mid-IR echo of a similar brightness to that seen \cite{jiang2019} in PS1-10adi would have been easily detectable, but no variability is seen in WISE data spanning $\approx1000$ days around the optical peak.

The AGN torus model can also explain \cite{jiang2019} two further features of PS1-10adi. The Fe II emission can occur from sublimation of the same dust responsible for the mid-IR echo (note that narrow Fe II can also arise from dense gas close to an accretion disk \cite{wevers2019}). Furthermore, the observed optical re-brightening $\sim2000$ days after the light curve peak (not typically seen in SNe) can arise when an outflow driven by the TDE eventually collides with the torus.

As well as the locations within their host galaxies differing, the properties of the galaxies themselves differ between SN2016aps and the nuclear events. The absolute magnitude of the SN2016aps host is $M_g\approx -16.9$\,mag, whereas the hosts of the other events are brighter by $>2.5-5$\,mag, i.e. a factor $10-100$ (note that these are also significantly brigher than most of the SLSN IIn hosts discussed in the previous section). The peak luminosity of PS1-10adi is consistent with the estimated Eddington luminosity of the associated AGN \cite{kankare2017}, and the other nuclear events are either less luminous or in brighter galaxies, i.e. all are likely radiating below the Eddington luminosity for the AGN in these galaxies. In comparison, the luminosity of SN2016aps would be approximately $10\times$ Eddington if it was associated with a supermassive black hole in such a faint galaxy, assuming typical scaling of the black hole mass with the galaxy mass \cite{kormendy2013}. Thus while all the nuclear events can be naturally accommodated within the context of black hole accretion, from TDEs or otherwise, SN2016aps is much more likely a SN.

The interacting SN model proposed for the nuclear transients can account for the dense CSM, either as a result of runaway mergers, an existing narrow-line region, and/or ionisation confinement in the AGN radiation field \cite{kankare2017}. However, dense CSM is not the only requirement; a large explosion energy $E_k > E_{\rm rad}$ is also needed to power the light curve. We are not aware of any reason why such explosions would occur preferentially near AGN. Most observed (SLSNe, long gamma-ray bursts) and theoretical (PISNe) SNe with large explosion energies favour low metallicity environments \cite{lunnan2014,schulze2018,yusof2013}, quite unlike the centres of massive galaxies. Moreover, the integrated radiation from PS1-10adi is somewhat uncomfortable for SN models, exceeding by a factor $\gtrsim2$ the maximum predicted emission for hydrogen rich SNe \cite{sukhbold2016}. We therefore conclude that extremely energetic nuclear transients are quite unlikely to be SNe, leaving SN2016aps as the most secure case of a SN radiating $>5\times10^{51}$ erg.

\subsection{Light curve models}

\textit{Models from the literature.} We estimate the mass and energetics of SN2016aps using analytic relations \cite{chevalier2011} for the interaction of SN ejecta with a dense wind. The total radiative energy released is given by
\begin{equation}
    E_{\rm rad} = 0.44\times10^{50} \kappa_{0.34}^{0.4} E_{51}^{1.2} M_{10}^{-0.6} D_*^{0.8}\,{\rm erg}
\end{equation}{}
where $\kappa_{0.34}$ is the opacity in units of $0.34$\,cm$^2$\,g$^{-1}$, $E_{51}$ the kinetic energy in units of $10^{51}$\,erg, $M_{10}$ is the ejected mass in units of $10$\,\M, and $D_*\equiv 1000 \dot{M}/v_w$ is the density parameter for pre-explosion mass-loss with rate $\dot{M}$ in \M\,yr$^{-1}$ and velocity $v_w$ in \kms. $E_{\rm rad}$ is obtained by integrating the bolometric light curve. The luminosity is given by
\begin{equation}
    L = 7.6\times10^{43} \kappa_{0.34}^{-0.6} E_{51}^{1.2} M_{10}^{-0.6} D_*^{-0.2}\,{\rm erg}\,{\rm s}^{-1}.
\end{equation}{}
We divide these two equations to eliminate $E_{51}$ and $M_{10}$, and assume $\kappa =0.34$\,cm$^2$\,g$^{-1}$ as appropriate for electron scattering in hydrogen-rich material. This gives the useful relation $D_*\propto E_{\rm rad}/L$. i.e., a flatter light curve (longer time-scale) indicates a higher density. We find $D_*=20.5$. This gives $\dot{M}\gtrsim 0.1-10$\,\M\,yr$^{-1}$ for $v_w=10-1000$\,\kms, fully consistent with our result in the main text and Figure 4 that used a different relation for the wind density \cite{chugai1994}. Putting these values back into either equation allows one to find $E_K^2/M_{\rm ej}$.

We then find the interaction radius by putting these values into the relation
\begin{equation}
    R_{\rm d} = 4.0\times10^{14} \kappa_{0.34}^{0.8} E_{51}^{0.4} M_{10}^{-0.2} D_*^{-0.2}\,{\rm cm},
\end{equation}{}
valid as long as the outer CSM radius is much greater than $R_d$. This gives $R_d=5.3\times10^{15}$\,cm, which is reassuringly consistent with the blackbody radius of the continuum emission (Extended Data Figure 1). This implies that most of the continuum emission comes from close to the contact discontinuity.

Finally, the shock velocity is found using
\begin{equation}
    R_{\rm d} = 5.7\times10^{14} \kappa_{0.34} D_* v_{\rm sh}\,{\rm cm},
\end{equation}{}
where $v_{\rm sh}$ is in units of $10^4$\,\kms. Putting in our earlier results gives $v_{\rm sh}\sim4600$\,\kms. Interestingly, this means that the transition to a steeper light curve at $\sim200$ days (Figure 4) corresponds to the doubling timescale of the shocked region.

We also compare to published SLSN IIn models from more realistic simulations \cite{dessart2015} in Figure 4. The data are reasonably consistent with models calculated for a CSM mass of 17.3\,\M\ and explosion energy between $3-10\times10^{51}$\,erg ($3-10$ times larger than a canonical SN). The ejecta mass in the model is 9.8\,\M; the sensitivity to this parameter was not explored in that study. However, we note that the steeper and earlier drop in the model luminosity compared to SN2016aps may be an indication that a larger mass is needed to match this event.

\textit{Bayesian light curve fit.} We fit a circumstellar interaction model to the observed UV and optical photometry using \textsc{mosfit}: the Modular Open Source Fitter for Transients \cite{guillochon2018}. This is a semi-analytic code employing a range of modules that can be linked together to produce model light curves of astronomical transients, and determine the best fitting model parameters through Bayesian analysis. The interaction model and its implementation in \textsc{mosfit} are described in a number of previous works \cite{chatzopoulos2012,chatzopoulos2013,villar2017}.

We first demonstrate that the model gives a reasonable match to the light curve using the parameters derived in the previous section. We take the lower limit on ejected mass and assume $M_{\rm ej}=52$\,\M, and the integrated shocked CSM mass \cite{chugai1994}, $M_{\rm CSM}=40$\,\M. We further assume that the observed photospheric radius corresponds to the contact discontinuity (i.e. the inner CSM radius), and the mass above this radius is set by our derived $D_*=20.5$ (with a corresponding wind profile for the CSM). We use $n=10$ and $\delta=1$ for the ejecta outer and inner density profiles, though are results are largely insensitive to these parameters. We set $v_{\rm ej}=10^{4}$\,\kms, larger than our derived shock velocity, but required to give a total energy $E_K=3.1\times10^{52}$\,erg and match the peak luminosity.  The result, shown in Extended Data Figure 2, gives a good match to the observations.

Next we free these parameters to find the best fit and Bayesian posteriors for our parameters. To sample the parameter space we used the affine-invariant ensemble method \cite{Goodman2010,Foreman2013}. We ran the Markov Chain with 100 walkers for 25,000 iterations, checking for convergence by ensuring that the Potential Scale Reduction Factor was $<1.2$ at the end of the run \cite{Brooks1998}. Our model has 7 free parameters: the masses of the star and CSM; the ejecta velocity; the inner radius of the CSM; the density at this inner radius; the time of explosion; and a white-noise term parameterising any unaccounted-for variance. We use the same priors as for SN2016iet \cite{gomez2019}, with a few modifications. We fix the opacity at $\kappa=0.34$\,cm$^2$\,g$^{-1}$, appropriate for electron scattering in hydrogen-rich matter, and the final continuum temperature at 6000\,K based on our photometry (Extended Data Figure 1). If left free, the temperature posteriors always converged to this value anyway, so we fixed it to speed up our model runs. We run one model for a shell-like (constant density) CSM, and one for a wind-like ($\rho\propto r^{-2}$) CSM, but otherwise use the same priors for both models.

To further reduce the number of free parameters, we assume 100\% efficiency in radiating the deposited energy. This efficiency follows that used in similar model fits \cite{chatzopoulos2013}; a lower efficiency would require a correspondingly larger explosion energy. The large efficiency is warranted as this model assumes an extremely optically thick interaction and therefore applies only in the limit of large masses. We note that for the analytic wind model \cite{chugai1994}, which does not require a large CSM mass, we used a lower efficiency of 50\%, also guided by previous work \cite{chevalier2012}. Assuming 100\% efficiency in that model would reduce the mass-loss curves in Figure 4 by a factor of 2, resulting in a total mass $M_{\rm CSM}\gtrsim 20$\,\M.

We obtain a similarly good fit for either a wind or shell CSM. The Watanabe-Akaike Information Criterion (WAIC) \cite{watanabe2010,gelman2014} is essentially indistinguishable between them, with WAIC\,$=147.0$ for the shell and 147.6 for the wind model.  We show the shell model in Figure 4 and the wind model in Extended Data Figure 2. The posterior probability densities of the free parameters is shown for both models in Extended Data Figure 3. In the wind model, some of the posteriors lie close to the upper bounds of the priors.

\subsection{Uncertainties in PISN rate estimates}

Both our PISN and PPISN rate estimates contain significant uncertainty, particularly in the $f_2$ parameter (Methods). Furthermore, the MESA models underpinning these calculations were computed at SMC metallicity. Retaining sufficient mass to reach the pair-unstable threshold depends on mass-loss rates that are highly sensitive to metallicty. Single star MESA models over a wider range in metallicity \cite{yusof2013} suggest that PISNe should not occur at all at solar metallicity. The host galaxy of SN2016aps likely has a metallicity intermediate between the LMC and SMC. At the higher metallicity of the LMC, PPISNe can occur for stars with initial masses $\gtrsim120$\,\M, while full PISNe require $\gtrsim300$\,\M. On the other hand, rapid rotation can lead to chemically homogeneous evolution and a larger core mass for a given initial mass, thus lowering these thresholds \cite{chatzopoulos2012b}. This also facilitates engine formation. Given these rather large uncertainties, all rate estimates should be considered indicative only.

However it is instructive to compare them to the observed rates of strongly interacting SNe. While literature estimates to date have been based on small numbers, the best current measurement \cite{quimby2013} of the SLSN II rate is $150^{+151}_{-82}$\,Gpc$^{-3}$\,yr$^{-1}$, corresponding to $\approx 3\times10^{-4}-1\times10^{-3}$ per CCSN. This is in broad agreement with the post-merger PPISN rate. The discovery of SN2016aps suggests that up to $\sim10\%$ of SLSNe IIn may exceed the energy budget of a typical SN; the estimated rates are consistent with such events being those that form magnetars.

We also note that an alternative engine could be fallback onto a central black hole remnant \cite{dexter2013}. Detailed rate estimates are not available for this model, but it seems to require relatively fine-tuned parameters in order to impact the observed light curve \cite{dexter2013}.

\bibliography{references}


\end{document}